\documentclass[letterpaper,aps,prd,twocolumn,tightenlines,preprintnumbers,nofootinbib,showkeys,superscriptaddress,longbibliography
]{revtex4-1}

\pdfoutput=1 

\usepackage[T1]{fontenc}

\usepackage{fullpage}
\usepackage{amsfonts}
\usepackage{amsmath}
\usepackage{slashed}
\usepackage{amssymb}
\usepackage{graphicx}
\usepackage{epic}
\usepackage{eepic}
\usepackage{epsfig}
\usepackage{latexsym}
\usepackage[dvipsnames]{xcolor}
\usepackage{float}
\usepackage{multirow}
\usepackage[breaklinks]{hyperref}
\usepackage{enumitem}
\hypersetup{colorlinks=true,citecolor=blue,linkcolor=blue,urlcolor=NavyBlue}
\usepackage[caption=false]{subfig}
\usepackage{natbib}
\usepackage{cancel}
\usepackage{relsize}
\usepackage[left=2cm,right=2cm,top=2cm,bottom=2cm]{geometry}
\usepackage{makecell}
\usepackage{soul}

\usepackage{bm}
\usepackage{subfig}

\usepackage{shorthand}
\usepackage{mciteplus}

\def\beq{\begin{equation}}
\def\eeq{\end{equation}}
\def\bea{\begin{eqnarray}}
\def\eea{\end{eqnarray}}

\def\h1{\ensuremath{h_1}}
\def\h2{\ensuremath{h_2}}

\newcommand{\hpm}{H^\pm}

\begin{document}
\linespread{1.02}

\title{Probing a 2HDM  Type-I light Higgs state via $H_{\rm SM} \to hh \to b\bar b\gamma \gamma$  at the LHC}



	\author{A. Arhrib}
	\email{aarhrib@uae.ac.ma}
	\affiliation{Abdelmalek Essaadi University, Faculty of Sciences and Techniques, B.P. 2117 T\'etouan, Tanger, Morocco.}
    \affiliation{Department of Physics and Center for Theory and Computation, National Tsing Hua University, Hsinchu, Taiwan 300.}

	\author{S. Moretti}
	\email{stefano.moretti@cern.ch}
	\affiliation{School of Physics and Astronomy, University of Southampton, Southampton, SO17 1BJ,\\ United Kingdom.}
	\affiliation{Department of Physics and Astronomy, Uppsala University, Box 516, SE-751 20 Uppsala, Sweden.}
	\affiliation{Particle Physics Department, Rutherford Appleton Laboratory, Chilton, Didcot, Oxon OX11 0QX, United Kingdom.}

	\author{S. Semlali}
	\email{souad.semlali@soton.ac.uk}
	\affiliation{School of Physics and Astronomy, University of Southampton, Southampton, SO17 1BJ,\\ United Kingdom.}
	\affiliation{Particle Physics Department, Rutherford Appleton Laboratory, Chilton, Didcot, Oxon OX11 0QX, United Kingdom.}
	
	\author{C.H.~Shepherd-Themistocleous}
	\email{claire.shepherd@stfc.ac.uk}
	\affiliation{Particle Physics Department, Rutherford Appleton Laboratory, Chilton, Didcot, Oxon OX11 0QX, United Kingdom.}
	\author{Y. Wang}
	\email{wangyan@imnu.edu.cn}	
	\affiliation{College of Physics and Electronic Information, Inner Mongolia Normal University, Hohhot 010022, PR China.}
	\affiliation{Inner Mongolia Key Laboratory for Physics and Chemistry of Functional Materials,Inner Mongolia Normal University, Hohhot, 010022, China.}
    \author{Q.S. Yan}
    \email{yanqishu@ucas.ac.cn}
	\affiliation{Center for Future High Energy Physics, Chinese Academy of Sciences, Beijing 100049, P.R. China.}
	\affiliation{School of Physics Sciences, University of Chinese Academy of Sciences, Beijing 100039, P.R. China.\\}

\date{\today}

\begin{abstract}
We study the discovery potential for a light Higgs boson via $gg \to H_{\text{SM-like}} \to hh \to b\bar{b}\gamma\gamma$ process at the Large Hadron Collider (LHC). Focusing on the 2-Higgs Doublet Model (2HDM) Type-I, which can accommodate light neutral Higgs states, of $\mathcal{O}(100)$\,GeV or less in mass, while agreeing with theoretical and up-to-date experimental constraints, we explore the feasibility of a light CP-even Higgs state $h$ via the largely unexplored final state $b\bar{b}\gamma\gamma$ at Run-3 of the LHC. We further propose a few Benchmark Points (BPs) for future searches.
\end{abstract}


\maketitle

\section{Introduction} \label{sec:intro}
The Large Hadron Collider (LHC) holds the promise to deepen our understanding of Electro-Weak Symmetry Breaking (EWSB) through its vast Higgs physics programme at Run3 as well as its future High-Luminosity (HL) upgrade. After the Higgs discovery in 2012 \cite{ATLAS:2012yve,CMS:2012qbp}, there has been an extensive, still ongoing at Run 3, effort to measure its properties with high precision~\cite{ATLAS:2022vkf,CMS:2022dwd}. This campaign has thus shifted the focus from initial observations to precision measurements. Ultimately, the upcoming HL-LHC datasets will consolidate such a new era of Higgs physics, enabling percent-level precision across most channels, thereby  significantly improving our understanding of the Higgs sector. 

Simultaneously, there has been a strong push to explore new physics beyond the minimal Higgs framework of the Standard Model (SM). One of the simplest and most extensively studied SM extensions is the 2-Higgs Doublet Model (2HDM), which introduces an extra Higgs doublet. This framework brings a rich phenomenology, including a wide range of Higgs-fermion interactions and additional new light and  heavy Higgs states. It also allows for exotic SM-like Higgs decays to two light (pseudo)scalars, $H_{\text{SM}} \to aa(hh)$, which then decay into pairs of SM  particles, offering a promising avenue to search for Beyond the SM (BSM) physics in various final states at the LHC. Present experimental data the from the ATLAS and CMS collaborations have, respectively, placed upper limits of 12\% \cite{ATLAS:2022vkf} and 16\% \cite{CMS:2022dwd} at 95\% C.L on the Branching Ratio (BR) of non-SM Higgs decays (${\rm{BR}}_\text{BSM}$).

Building on these recent experimental searches, we aim to probe the significantly unexplored $b\bar{b}\gamma\gamma$ final state in searches for $H\to aa(hh)$ cascade signals within the sub-62 GeV mass range. Currently, no upper limit exists on  ${{\rm{BR}}}(H \to aa \to b\bar{b}\gamma\gamma)$ for $m_a\in [10,~62]~\text{GeV}$. However, indirect constraints arise from searches for $H \to aa \to \gamma\gamma\gamma\gamma$. The CMS collaboration reported upper limits on $\sigma(pp \to H)\times {{\rm{BR}}}(H \to aa \to \gamma\gamma\gamma\gamma)$, ranging from 0.80 fb for $m_a=15~\text{GeV}$ to 0.26 fb for $m_a = 62~\text{GeV}$~\cite{CMS:2022xxa}, while the ATLAS collaboration  established further limits on  ${{\rm{BR}}}(H\to aa \to \gamma\gamma\gamma\gamma)$, ranging from $10^{-5}$ to 0.03, using their full Run 2 dataset with an integrated luminosity of $140~\text{fb}^{-1}$~\cite{ATLAS:2023ian}. A Complementary analysis from ATLAS Run 1 searches for $H \to aa \to \gamma\gamma jj$~\cite{ATLAS:2018jnf} placed additional indirect limits on ${{\rm{BR}}}(H \to aa \to b\bar{b}\gamma\gamma)$. Combining these limits with ${\rm{BR_{BSM}}} < 0.12$ enables setting indirect bounds on  ${{\rm{BR}}}(H \to aa(hh) \to b\bar{b}\gamma\gamma)$. It is noteworthy that processes featuring photon pairs in their final states, such as $H \to \gamma\gamma\gamma\gamma,~\gamma\gamma gg(jj)$, are particularly relevant in models where the light Higgs state decays predominantly to photons and has suppressed fermionic decays (an "almost fermiophobic" scenario), which, in turn, leads to enhanced bosonic decay rates in the  2HDM Type-I~\cite{Akeroyd:1995hg,Akeroyd:1998ui,Brucher:1999tx,Akeroyd:2003xi,Arhrib:2008pw,Arhrib:2017uon,Kim:2022nmm}. 

In this letter, we cover the final state $b\bar{b}\gamma\gamma$ via the process $\sigma(gg \to H_{\text{SM}})\times {{\rm{BR}}}(H_{\text{SM}} \to hh \to b\bar{b}\gamma\gamma)$. This study builds on our previous work~\cite{Arhrib:2023apw}, where we explored the sensitivity to the  2HDM Type-I parameter space using  $H_{\text{SM}} \to hh \to b\bar{b}\tau\tau$. By including the $b\bar{b}\gamma\gamma$ final state, we aim to provide a complementary test of the 2HDM framework and to broaden our understanding of potential BSM physics in exotic Higgs decays. Through detailed Monte Carlo (MC) simulations, we assess the feasibility of detecting the process $gg \to H_{\text{SM}} \to hh \to b\bar{b}\gamma\gamma$, during LHC Run 3 (and beyond). Our approach focuses on the  distinctive final state in order to identify optimised selection criteria that enhance signal-to-background discrimination and effectively isolate this signature from various background processes, i.e., $t\bar{t}H$, $ZH$, $b\bar{b}H$. With this study, we aim  to further shed light on the 2HDM Type-I parameter space and explore the possibility of confirming or excluding such a scenario through future LHC data.

This letter is organised as follows. In Sect.~\ref{sec:model}, we provide a brief overview of the 2HDM framework, followed by our simulation setup and analysis in Sect.~\ref{sec:numerical},   we  then conclude in Sect.~\ref{sec:summa}.


\section{The 2HDM Type-I}
\label{sec:model}

\subsection{Higgs potential}
\label{subsec:THDM}

The most general renormalisable Higgs  potential of a CP-conserving 2HDM, with a softly broken $Z_2$ symmetry, is given by~\cite{Branco:2011iw}:
\begin{equation}
\begin{split}
\mathcal{V} &= m_{11}^2\Phi_1^\dagger\Phi_1+ m_{22}^2\Phi_2^\dagger\Phi_2
-[m_{12}^2\Phi_1^\dagger\Phi_2+ \, \text{h.c.} ] \\
& +\frac{\lambda_1}{2}(\Phi_1^\dagger\Phi_1)^2
+\frac{\lambda_2}{2}(\Phi_2^\dagger\Phi_2)^2
+\lambda_3(\Phi_1^\dagger\Phi_1)(\Phi_2^\dagger\Phi_2)\\
&+\lambda_4(\Phi_1^\dagger\Phi_2)(\Phi_2^\dagger\Phi_1) 
+\left[\frac{\lambda_5}{2}(\Phi_1^\dagger\Phi_2)^2
+\,\text{h.c.}\right]\,.
\label{eq:2hdmpot}
\end{split}
\end{equation}
All the squared mass parameters, $m^2_{11},~m^2_{22}$ and $m^2_{12}$, along with the (dimensionless) quartic couplings $\lambda_{1-5}$ are assumed to be real-valued. Using the potential minimisation conditions, $m^2_{11}$ and $m^2_{22}$ can be replaced by the Vacuum Expectation Values (VEVs) $v_{1,2}$. At tree level, the quartic couplings  $\lambda_{1-5}$ can be expressed in terms of the four (physical) Higgs boson masses, the mixing angle $\alpha$ between the CP-even scalars and the angle $\beta$ which is defined through $\tan\beta = v_2/v_1$. The Higgs sector can be then described by 7 independent parameters:
\begin{eqnarray}
m_{H},~m_{h},~m_{H^{\pm}},~m_{A},~\alpha,~\beta~\text{and} \ m_{12}^2.
\label{eq:param} 
\end{eqnarray}

The presence of an extra Higgs doublet in the Yukawa sector introduces unwanted Flavour Changing Neutral Currents (FCNCs). By imposing a $Z_2$ symmetry, these FCNCs can be suppressed, resulting in four main types of 2HDM, where each is defined by how the Higgs doublets couple to fermions. Here, we focus on the Type-I realisation of the 2HDM, where only one of the Higgs doublets couples to all fermions. The neutral Higgs-fermions couplings, $\kappa_{\phi f\bar{f}}$ with $\phi=h,~H$, are given by \cite{Branco:2011iw}: 
\begin{eqnarray}
\kappa_{hf\bar{f}} = \cos\alpha/\sin\beta,\  \ \    \ \kappa_{Hf\bar{f}} = \sin\alpha/\sin\beta.
\end{eqnarray}

\subsection{Theoretical and  experimental constraints}
The 2HDM parameter space is tested against various constraints. On the theoretical side, all Higgs potential parameters must satisfy unitarity, vacuum stability and
perturbativity conditions. This has been checked using the public code \texttt{2HDMC-1.8.0}~\cite{Eriksson:2009ws}. 
On the experimental side, compatibility with EW Precision Observables (EWPOs) is ensured by requiring the computed $S$, $T$ and $U$ values from \texttt{2HDMC-1.8.0} to lie within 2$\sigma$ of the SM fit from~\cite{ParticleDataGroup:2020ssz}, while fully accounting for the correlations between the three parameters. Furthermore, constraints from $B$-physics observables, Higgs searches at lepton and hadron colliders as well as Higgs signal measurements are  checked by \texttt{SuperIso-v4.1}~\cite{Mahmoudi:2008tp} and \texttt{HiggsTools-v1.2}~\cite{Bahl:2022igd}, respectively.

\section{Numerical results}\label{sec:numerical}
\subsection{Parameter space}
\label{sec:Signal}
Focusing on the inverted hierarchy, where the heaviest Higgs state is identified as the observed 125 GeV object, we conducted a random scan over the following ranges:
\begin{center}
	$m_h$: $[15,~60]$\,GeV\,,~~$m_A$: $[62,~100]$\,GeV\,,\\
	$m_{\hpm}$: $[100,~200]$\,GeV\,,~~$s_{\beta - \alpha}$: $[-0.25,~-0.05]$\,,\\
	~~$m_{12}^2$: 0 -- $m_h^2\cos\beta \sin \beta$\,,~~$t_\beta$: 2 -- 25\,. \\
\end{center} 
 
\begin{figure}[h!]
	\includegraphics[scale=0.4]{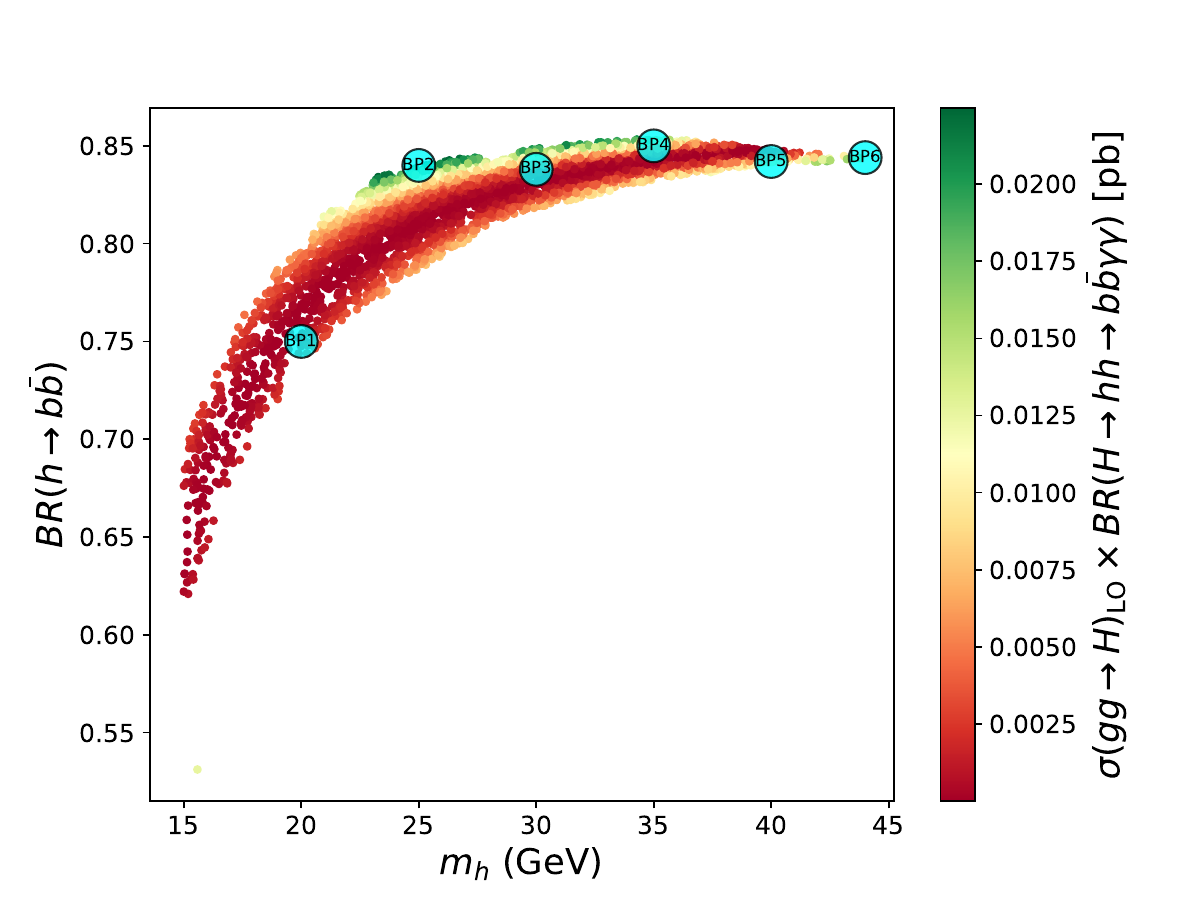}
	\caption{${\rm{BR}}(h\to b\bar{b})$ as a function of $m_h$ vs. $\sigma(gg \to H \to hh \to b\bar{b}\gamma\gamma)$. BPs are marked in cyan colour.}
	\label{fig1}
\end{figure}

The scanned parameter points, which satisfy both theoretical and experimental constraints, are presented in Fig.~\ref{fig1}. This figure illustrates the BR of $ h \to b\bar{b}$ as a function of the light Higgs mass ($m_h$), compared against the overall cross-section for the process $gg \to H \to hh \to b\bar{b}\gamma\gamma$. Here, we focus on the main production mechanism of the SM-like Higgs boson ($H$) at the LHC, which is gluon-gluon fusion (ggF). The Higgs production cross section, $\sigma(gg \to H)$, is computed at Leading Order (LO) by \texttt{Sushi}~\cite{Harlander:2016hcx} at the center-of-mass energy of 14 TeV, whilst the QCD corrections are considered through a $K$-factor~\cite{Cepeda:2019klc}.

The $h\ \to b\bar{b}$ decay has the largest BR, of $\sim$ 60-85 \%, due to enhanced light Higgs couplings to fermions, while the decay rate of $h \to \gamma\gamma$ is suppressed, exhibiting a BR below 12\% that decreases to approximately 3\% for $m_h \gtrsim 30$ GeV. The LO cross section reaches up to 0.02 pb when ${{\rm{BR}}}(H \to hh)$ and ${{\rm{BR}}}(h\to b\bar{b})$ are at their maximum. Notably, a significant region of the parameter space has been excluded by the recent CMS search for $H \to aa \to \gamma\gamma\gamma\gamma$~\cite{CMS:2022xxa} at $\sqrt{s} = 13$ TeV, with an integrated luminosity of 132 $\text{fb}^{-1}$. Additionally, further constraints on the parameter space arise from Higgs signal requirements, which restrict here ${{\rm{BR}}}(H \to hh)$ to below 4\%.

From the parameter points that successfully passed all constraints, several Benchmark Points (BP) were identified to perform a MC simulation. The selected signal samples correspond to a scalar boson mass, $m_h$, ranging from 20 to 45 GeV in approximately 5 GeV steps to ensure a consistent and systematic exploration of the parameter space as shown in Tab.~\ref{tab1}.

\begin{table}[h!]
	\begin{center}
		\setlength{\extrarowheight}{0.02cm} 
		\resizebox{0.42\textwidth}{!}{
			\begin{tabular}{c|c|c|c|c|c|c} \hline\hline
				BP & $\text{BP}_{1}$ & $\text{BP}_{2}$ & $\text{BP}_{3}$ & $\text{BP}_{4}$ & $\text{BP}_{5}$ & $\text{BP}_{6}$ \\
				\hline \hline
				$m_h$  & 20 & 25 & 30 & 35 &40 & 44 \\
				\hline 
				 $\sin(\beta -\alpha)$  & -0.12 & -0.07& -0.05& -0.06 & -0.077  & -0.11\\
				 \hline
				  $\tan\beta$  & 8 & 13&20 & 17& 14 & 10 \\
				  \hline 
				  $m_A$ & 89 & 86 & 87& 90 & 94 & 99 \\
				  \hline 
				   $m_{H^\pm}$ & 125 & 119& 109 & 108 &  116& 145 \\
				   \hline \hline
				   \multicolumn{7}{c}{BRs [\%]} \\
				   \hline\hline
				    ${\rm{BR}}( H \to hh)$ &0.35 &2.4 & 3.4&3 & 1.33 &  2.14 \\
				    				   \hline	
				   ${\rm{BR}}( h \to b\bar{b})$ & 79&83 &84 &85.2 & 84.2 & 84.3 \\
				   \hline
				   ${\rm{BR}}( h \to \gamma\gamma)$ &6.4 &2.9 &2.3 &2.08 & 3.17 &  2.85 \\				  			   
				   \hline \hline
				   \multicolumn{7}{c}{cross section [fb]}  \\
				   \hline\hline
				    $\sigma_{b\bar{b}\gamma\gamma}^{\text{LO}}$ &6.10 & 20.0 & 22.0 & 17.7  & 12.0  & 17.4\\ 	
				 \hline  \hline
			\end{tabular}
		}
	\end{center}
    \vspace*{-0.3cm}
	\caption {Parameters for ${\text{BP}}_{1-6}$. All masses are in GeV, with $m_H = 125~\text{GeV}$. The LO cross sections $\sigma_{b\bar{b}\gamma\gamma} = \sigma(gg \to H) \times {\rm BR}(H \to hh \to b\bar{b}\gamma\gamma)$ for our BPs are given for the collision energy $\sqrt{s} = 14~\text{TeV}$. The $K$-factor mentioned above will be used later to compute the final significances.}
	\label{tab1}
\end{table}

\subsection{Signal vs. background analysis}
The generation of parton-level events for both signal and background processes is performed using \texttt{MadGraph-v3.4.2}~\cite{Alwall:2014hca}. These events are subsequently passed to \texttt{PYTHIA8}~\cite{Sjostrand:2006za} for parton showering, fragmentation, hadronisation and the decays of heavy-flavour particles. Detector effects are then simulated using \texttt{Delphes-3.5.0}~\cite{deFavereau:2013fsa}, with a standard CMS card. Finally, events selection and kinematic analysis are carried out using \texttt{MadAnalysis 5} \cite{Conte:2012fm}. The following cuts are applied at parton level for an efficient generation of signal and background events ($j$ here refers to a quark or gluon):
\begin{equation}
p_T(\gamma[b,j])>5[10]~\text{GeV}, |\eta(\gamma,b, j)|<3.5,~\Delta R(\gamma,b,j)> 0.3.
\label{cut}
\end{equation}
The major SM background processes are the production of a single Higgs boson in association with a top-quark pair ($t\bar{t}H$), $Z$ boson ($ZH$)  and bottom-quark pair $(b\bar{b}H)$, followed by $ H \to \gamma \gamma$. Tab.~\ref{tab2} summarises the LO cross sections for the different background processes, generated at $\sqrt{s}=14$ TeV\footnote{{Additional processes such as $pp \to t\bar{t}h$, $Zh$, $b\bar{b}h$ with $h$ being the light Higgs state were also generated. However, their contributions were found to be negligibly small.}}.
\begin{table}[h!]
	\begin{center}
			\setlength{\extrarowheight}{0.05cm} 
		\resizebox{0.4\textwidth}{!}{
			\begin{tabular}{|c|c|} \hline\hline
				Process  & $\sigma$ [fb] \\ 
				\hline
				$pp \to t\bar{t}H,~t\to bW^+,~\bar{t}\to\bar{b} W^-,~H(\to \gamma\gamma)$ &   0.61\\
				\hline 
				$gg \to t\bar{t}H,~t\to bW^+,~\bar{t}\to\bar{b} W^-,~H(\to \gamma\gamma)$ &  0.46  \\
				\hline 
				$pp \to Z(\to jj)H(\to \gamma\gamma),~j=j,~b,~\bar{b}$ &  0.70  \\
				\hline 
				$pp \to Z(\to b\bar{b})H(\to \gamma\gamma)$ &  0.15 \\
				\hline 
				$pp \to b\bar{b}H(\to \gamma\gamma)$ &  0.76\\
				\hline \hline
		\end{tabular}}
	\end{center}
	\caption {The LO background cross sections at parton level for the collision energy $\sqrt{s}=14$ TeV. The $K$-factors  mentioned in the text are not included here but will be discussed later.}
	\label{tab2}
\end{table}

After detector simulation, we require the presence of two final state $b$-jets and two photons in each event that satisfy the following criteria:
\begin{eqnarray}
p_T(b_1/b_2) &>& 20/20~\text{GeV},~p_T(\gamma_1/\gamma_2)> 22/14~\text{GeV}, \nonumber \\
|\eta(b)|&<& 2.5,~|\eta(\gamma)|<1.5
\end{eqnarray}

In searching for light Higgs bosons in the $\gamma\gamma$ channel, the CMS (high-level) di-photon trigger relies on the presence of two photons where the leading has $p_T$ threshold of 30 GeV and the subleading one has a $p_T>18$ GeV~\cite{CMS:2022xxa}. Given the presence of soft final states originating from light Higgs decays, we opted to refine our trigger strategy by focusing exclusively on barrel photons, rather than using an inclusive trigger. For barrel-only photon pairs (i.e., photons within $|\eta|<1.5$), the trigger thresholds are approximately 22 (14) GeV for the leading (subleading) photon. To further optimise photon candidate identification by rejecting fake photon candidates, we require that the energy deposited in the Hadronic CALorimeter (HCAL) does not exceed 10\% of the energy recorded in the Electromagnetic CALorimeter (ECAL). Furthermore, misidentified photons are rejected using the isolation variables $I_{\pm}$,  $I_0$  and $I_{\gamma}$, which quantify the calorimetric deposits from charged hadrons, neutral hadrons and photons, respectively, within a cone of radius $\Delta R =0.3$ around the photon candidate~\cite{CMS:2015myp}.


For jets clustering, we adopt anti-$k_T$ algorithm~\cite{Cacciari:2008gp} with a jet radius parameter $\Delta R =0.4$ and $p_T^{min} = 20$ GeV (for both light- and $b$-jets). Requiring exactly two $b$-tagged jets in each event with a transverse momentum of $p_T(b) > 20$ GeV imposes a tight condition due to the low  $p_T$ values and the limited efficiency of $b$-tagging. This is especially challenging in scenarios with low Higgs masses below 60 GeV, where the resulting $b$-jets typically have lower $p_T$ values. In such cases, the efficiency of $b$-tagging decreases substantially, as $b$-tagging algorithms are optimised for higher $p_T$ jets~\cite{Cagnotta:2022hbi}. 
In our previous analysis~\cite{Arhrib:2023apw}, we explored how such an efficiency varies with different $p_T$ thresholds. We observed that, while alternative pre-selection thresholds (such as $p_T(b_1/b_2) > 20/15$ or $15/10$ GeV) retain more signal events, the 20/20 GeV threshold ultimately offers better event significance, despite a higher loss in signal events. Thus, we require the 20/20 GeV threshold for the current analysis.

To enhance the sensitivity of our analysis and improve the signal-to-background ratio, we examine the reconstructed mass of the entire system ($m_{H}\equiv m_{b\bar b\gamma\gamma}$) alongside other relevant variables, such as the invariant mass of the $b$-jets ($m_{b\bar{b}}$) and  di-photon system ($m_{\gamma\gamma}$). In signal events, these variables are expected to be low as the objects originate from a light Higgs boson, $h$. Conversely, background events generally display higher values, as they do not stem from the decay of a light resonance. This distinction enables an effective separation of signal from background.

\begin{figure}[h!]
	\includegraphics[scale=0.4]{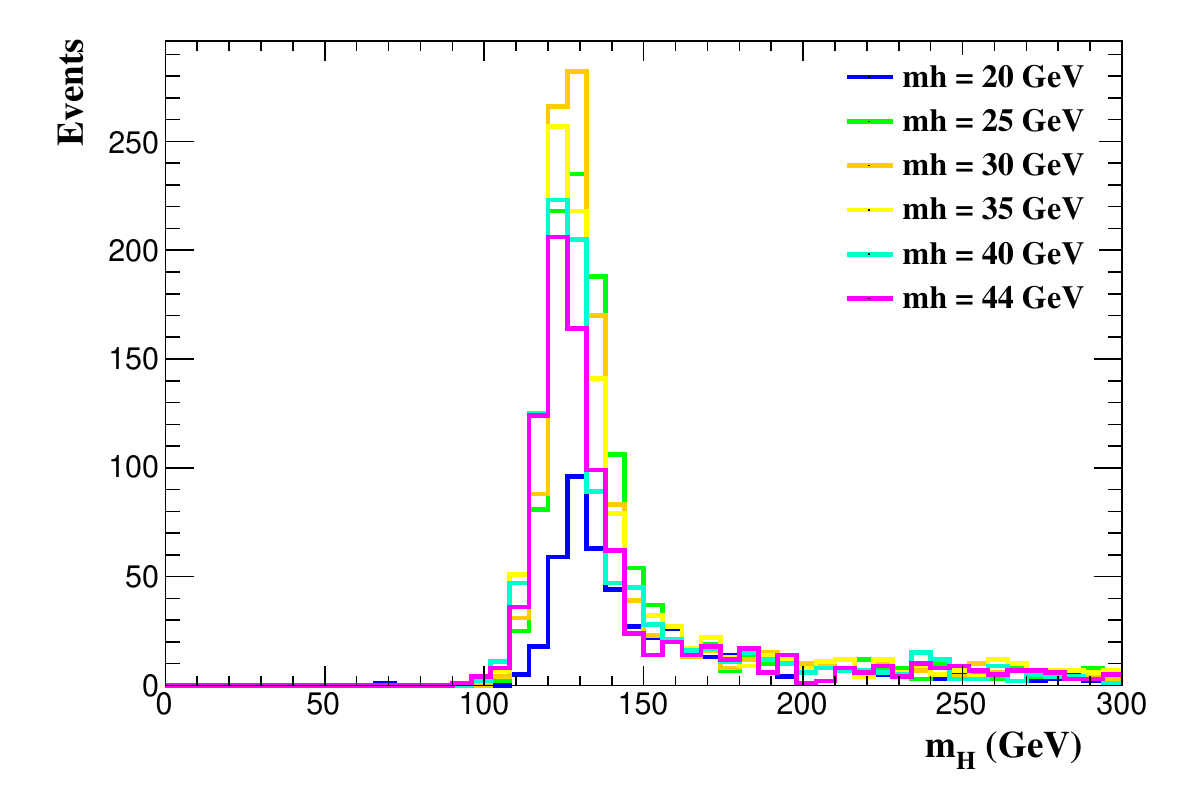}
	\caption{The distributions of $m_H$ for different BPs are shown at detector level.}
	\label{fig2}
\end{figure}

In Fig.~\ref{fig2}, the reconstructed SM-like Higgs boson mass distribution is consistent with the expected Higgs signal. Applying a kinematic cut on $m_H$ would then significantly improve the isolation of signal events from background noise, given that signal events, originating from Higgs boson decays, produce a sharp mass peak around 125 GeV, whereas background events, arising from $t\overline{t}$ and $ZH$, would display a broader distribution at higher masses.

In Tab.~\ref{tab3}, we present the event rates of the signal after applying our selection criteria. Following the extraction of events containing two photons and two $b$-tagged jets, we impose an additional requirement, that the invariant masses of the $b$-jet  ($m_{b\bar{b}}$) and photon ($m_{\gamma\gamma}$) pair must be less than $m_H/2$. This criterion captures the SM-like Higgs boson decaying into two light Higgs bosons, $H\to hh$, in the signal events.
To further enhance the separation between signal and background, we introduce a new variable, $\Delta m_h = (m_{b\bar{b}} - m_{\gamma\gamma}) / m_{\gamma\gamma}$, which measures the relative difference between $m_{b\bar{b}}$ and $m_{\gamma\gamma}$. Applying a cut of $\Delta m_h < 0.25$ effectively suppresses background events, as there is no intrinsic correlation between $m_{b\bar{b}}$ and $m_{\gamma\gamma}$ in the latter, unlike in signal events, where both $b\overline{b}$ and $\gamma\gamma$ pairs originate from light Higgs bosons of the same mass. 
Finally, we set a cut on the reconstructed mass of the entire system, consistent with the expectation that the energies of these reconstructed light Higgs bosons  are capped at a mass of 125 GeV.

\vspace*{0.8cm}
\begin{table}[h!]
	\begin{center}
	\setlength{\extrarowheight}{0.02cm} 
	\resizebox{0.45\textwidth}{!}{
		\begin{tabular}{|c||c|c|c|c|c|c|} \hline \hline
			BP &  $\text{BP}_{1}$ &  $\text{BP}_{2}$& $\text{BP}_{3}$&  $\text{BP}_{4}$  &  $\text{BP}_{5}$ &  $\text{BP}_{6}$\\ 
			\hline 
			$m_h$ (GeV) & 20 & 25 & 30 & 35 & 40 & 44 \\
			\hline
			NoE &  823.5  &  2700      &  2970.000    &  2591      &   1980  & 3027 \\
			\hline
			$\gamma\gamma$& 172.14   &   593    &  635.10    & 518.87  & 402.19   & 600.67 \\
			\hline 
			2 $b$-jets& 4.75   &  33.39     &   38.10   &   33.06   & 21.879    & 30.45 \\
			\hline 
			$m_{bb}<62$ GeV&  3.12  &  27.02     &   30.41   &  25.78  & 16.03 & 21.98 \\
			\hline 
			$m_{\gamma\gamma}<62$ GeV& 3.12   &  27.00     &   30.35   &   25.73 & 15.97 & 21.92  \\
			\hline 
			$\Delta m_h<0.25$& 1.92    &   20.68   &    25.33       &    21.09  & 13.26  & 18.8\\
			\hline 
			$m_{H}<150.0$ GeV& 1.91   &  20.57     &   25.18        &   20.75 &  13.10  & 18.55\\
			\hline \hline 
		\end{tabular}
	}
    	\end{center}	
	\caption {Signal events rates after applying basic and mass cuts for $\mathcal{L} = 300~\text{fb}^{-1}$.}
		\label{tab3}
\end{table}

Tab.~\ref{tab4} presents the cutflow results for the dominant background processes. Clearly, the mass observables, i.e., $m_{b\bar{b}}$, $m_{\gamma\gamma}$ and $\Delta m_h$, can greatly suppress the background events, thereby enhancing the sensitivity to our signal process\footnote{In the mass range where $m_A > m_h + m_Z$, the decay process $gg \to A \to hZ \to  b\bar{b} \gamma\gamma$ can mimic the signal. However, after performing an additional scan across the region $m_A \in [60, 200]$ GeV, we found that applying the specified kinematic cuts would effectively suppress this background, particularly the cuts on $m_{bb}$ and $\Delta m_h$, as in our study, the $b\bar{b}$ pair originates from a light Higgs resonance, in contrast to the process where $Z \to b\bar{b}$ and no correlation exists between $m_{b\bar{b}}$ and $m_{\gamma\gamma}$. Moreover, requiring a cut on the reconstructed mass of the full system further reduces noise from this background. In this analysis, we only present results specifically for $m_A \in [60, 90]$ GeV.}.
It is also worth noting that the QCD corrections for the background processes are not included here. Given that our selection criteria already substantially suppress the background events, the QCD corrections to the background would have a negligible effect on the overall sensitivity of our analysis. With background contributions being so limited after selection cuts, we safely neglect them when calculating the significance of the signal.

\begin{table}[h!]
		\begin{center}
	\setlength{\extrarowheight}{0.02cm} 
     \resizebox{0.47\textwidth}{!}{
		\begin{tabular}{|c|| c|c|c|c|c|} \hline\hline
			Process & $Z(\to bb)H$ & $Z(\to jj)H$ & $pp \to t\bar{t}H$ & $gg \to t\bar{t}H$ & $pp \to bbH$  \\ 
			\hline 
			NoE   & 45.00 & 210 & 183 & 138& 228.0\\
			\hline
			$\gamma\gamma$& 12.38 & 57.52 & 49.75 & 37.51& 72.24\\
			\hline 
			2 $b$-jets& 3.05 &3.11 &19.29 & 14.66&8.51 \\
			\hline 
			$m_{bb}<62$ GeV&~  0.44 & 0.31 & 1.26&0.91& 1.53 \\
			\hline 
			$m_{\gamma\gamma}<62$ GeV&0.002 & - & 0.002 & 0.003& 0.005 \\
			\hline 
			$\Delta m<0.25$& - & - &- &-& 0.005\\
			\hline 
			$m_{H}<150.0$ GeV& - & - &- &-&- \\
			\hline \hline 
		\end{tabular}
	}
	  	\end{center}
	\caption {Backgrounds events rates after applying basic and mass cuts for $\mathcal{L} = 300~\text{fb}^{-1}$.}
		\label{tab4}
\end{table}

Tab.~\ref{tab5} presents the significance ($\Sigma$) values for BP, where $\Sigma = N_S/\sqrt{N_S + N_B} \approx N_S/\sqrt{N_S} \approx \sqrt{N_S}$, where $N_{S}$ and $N_{B}$ indicate the event rates of the signal and backgrounds, respectively. Higher order corrections for signal are quantified through $K$-factor ($K \sim 2.4 -2.5$)~\cite{Cepeda:2019klc}. One can read from the table that five of our BPs are within the discovery reach at Run 3 of the LHC, with a significance above $5\sigma$. Needless to say, significances at the HL-LHC would be further increased by a factor approximately $\sqrt{10}$, thereby enabling access also to the remaining BP. 
 
\begin{table}[h!]
	\begin{center}
			\setlength{\extrarowheight}{0.1cm} 
		\resizebox{0.45\textwidth}{!}{
			\begin{tabular}{|c||c|c|c|c|c|c|} \hline\hline
				Benchmark Points (BP) & $\text{BP}_{1}$ & $\text{BP}_{2}$ & $\text{BP}_{3}$ & $\text{BP}_{4}$ & $\text{BP}_{5}$ & $\text{BP}_{6}$  \\ 
				\hline 
				$m_h$ (GeV) & 20 & 25 & 30 & 35 & 40 & 44 \\
				\hline
				Significance ($\Sigma$) &2.18  & 7.07&7.9& 7.2 &5.72 & 6.8\\
				\hline \hline 
		\end{tabular}}
	\end{center}
	\caption {Significances for our signal with $\sqrt{s} = 14$ TeV and $\mathcal{L}=300~\text{fb}^{-1}$. }
		\label{tab5}
\end{table}


\section{Conclusions}
\label{sec:summa}
This letter explored the parameter space of the 2HDM Type-I, specifically, within the framework of the so-called inverted mass hierarchy. In this scenario, the heaviest CP-even Higgs state, $H$, is identified as the SM-like Higgs boson ($H_{\text{SM}}$), while $h$ corresponds to a lighter Higgs state. After taking into account all available theoretical and experimental constraints from standard tools, alongside recent limits from searches for exotic Higgs decays into two lighter (pseudo)scalars, we investigated the final state $b\bar{b}\gamma\gamma$ to search for light Higgs bosons in the decay channel, $H_{\text{SM}} \to hh$, where the main production mechanism of the SM-like Higgs is gluon-fusion (ggF). 

We have not included some major SM background processes in our analysis, like the irreducible processes $pp\to bb\gamma\gamma$, and reducible processes $pp\to bb jj$, $pp\to j j j j$, and $pp \to jj \gamma \gamma$. These processes were omitted based on considerations of b-tagging efficiency and photon mistagging rates. Using these efficiencies and mistagging probabilities, we estimated their contributions to be negligible compared to the dominant backgrounds already considered in our study. Furthermore, generating the data sample of these processes demands huge computing resources, since such SM processes typically have large cross sections and they might contribute to the background via mistagging jets as photons or mistagging light jets as b-jets. Estimating such backgrounds demands reliable theoretical calculation and appropriate detector simulations. To circumvent these challenges, experimental groups use the data-driven approaches. For example, these background processes have been taken into account in realistic data analysis as demonstrated in \cite{ATLAS:2018jnf,CMS:2022xxa,ATLAS:2023ian}.

Through a MC analysis of the signal and background processes, exploiting several BPs representative of the parameter space of the chosen BSM framework, we have demonstrated that there is a sensitivity to the overall 2HDM Type-I signal $gg \to H_{\text{SM}} \to hh \to b\bar{b}\gamma\gamma$ already at the ongoing Run 3 of the LHC, with full coverage eventually being attained at the HL-LHC. Therefore, this analysis reveals that the 2HDM Type-I offers a viable search avenue for BSM physics with an enlarged Higgs sector with respect to the SM.

\begin{acknowledgments} 
We would like to thank Sam Harper for his invaluable input and discussions around the trigger analysis. SM is supported in part through the NExT Institute and the STFC Consolidated Grant   ST/X000583/1.
CHS-T(SS) is supported in part(full) through the NExT Institute. SS acknowledges the use of the IRIDIS High Performance Computing Facility, and associated support services at the University of Southampton, in the completion of this work. YW’s work is supported by the Natural Science Foundation of China Grant No. 12275143, the Inner Mongolia Science Foundation Grant No. 2020BS01013 and the Fundamental Research Funds for the Inner Mongolia Normal University Grant No. 2022JBQN080.
QSY is supported by the Natural Science Foundation of China under the Grants No. 11875260 and No. 12275143.

\end{acknowledgments} 

%

\twocolumngrid

\bibliography{bibio}

\end{document}